\documentclass[aps,pre,twocolumn,superscriptaddress]{revtex4-1}
\usepackage{amssymb}
\usepackage{amsmath,bm}
\usepackage{graphicx,color}
\usepackage{bbm}
\usepackage[bottom]{footmisc}
\usepackage{multirow}
\usepackage{xcolor}
\usepackage{xpatch} % <<<<<<<<<<<<<
\makeatletter   
\usepackage{xpatch} % <<<<<<<<<<<<<

\newcommand{\B}[1]{{\bm{#1}}}%% Bold Roman & Greek Lower & Upper Case

\newcommand{\rin}{r_\text{in}}
\newcommand{\rout}{r_\text{out}}

\usepackage{float}

\begin{document}
\title{Anomalous Elasticity in Classical Glass-formers}
\author{Avanish Kumar}
\affiliation{Dept. of Chemical Physics, The Weizmann Institute of Science, Rehovot 76100, Israel}
\author{Michael Moshe}
\affiliation{Racah Institute of Physics, The Hebrew University of Jerusalem, Jerusalem, Israel 9190}
\author{Itamar Procaccia$^*$} 
\affiliation{Dept. of Chemical Physics, The Weizmann Institute of Science, Rehovot 76100, Israel, $^*$Center for OPTical IMagery Analysis and Learning, Northwestern Polytechnical University, Xi'an, 710072 China}
\author{Murari Singh}
\affiliation{ McKetta Department of Chemical Engineering, University of Texas at Austin, Austin, Texas 78712}

\begin{abstract}
Amorphous solids under mechanical strains are prone to plastic responses. Recent work showed that in amorphous granular system these plastic events, that are typically quadrupolar in nature, can screen the elastic response. When the density of the quadrupoles is high, the gradients of the quadrupole field act as emergent dipole sources, leading to qualitative changes in the mechanical response, as seen for example in the displacement field. In this paper we examine the effect of screening in classical glass formers. These are made of point particles that interact via binary forces. Both inverse power law forces and Lennard-Jones interactions are examined, and it is shown that in both cases the elastic response can be strongly screened, in agreement with the novel theory. The degree of deviation from classical elasticity theory is parameterized by a proposed new measure that is shown to have a functional dependence of on the amount of energy lost to plastic responses. 
\end{abstract}
\maketitle

\section{Introduction}

The term ``amorphous solids" refer to glasses, foams, emulsions, colloidal suspensions, granular media etc., when their constituents are packed densely enough either by mechanical forces or due to low temperatures. They are amorphous since they do not have long range order, and they are solids since they have shear and bulk moduli that appear to dictate their response to external strains \cite{89Lutsko}. It is therefore tempting to assume that classical elasticity theory should apply to such solids, and that, given the knowledge of the elastic moduli, one can predict the displacement field that is associated with any mechanical strain \cite{Landau}. In fact, recent research indicates that this may not be the case. Firstly, in contradistinction with perfect crystalline solids, in amorphous solids {\em plastic responses} appear instantaneously for any amount of strain \cite{10KLP}. Secondly, it was shown that {\em nonlinear}  elastic
moduli of amorphous solids have unbounded sample-to-sample fluctuations in the thermodynamic limit \cite{11HKLP}. Moreover, at least in frictional amorphous solids, one observes stress correlations that are not consistent with classical elasticity theory \cite{21LMPR,21LMPRWZ}. Under large strains amorphous solids can show yield regimes in which stress does not increase when strain does, indicating clearly that elasticity theory in its present classical form cannot describe satisfactorily important examples of mechanical responses of amorphous solids. Indeed, recent theory indicates that plastic events that abound in amorphous solids can screen the mechanical fields, resulting in possible strong renormalization of the elastic moduli and in mechanical

%%%%%%%%%%%%%%%%%%%%%%%%%%%%%%%%%%%%%%%%%%%%%%%%%%%%%%%%%%%
\begin{figure}[htb!]
	\includegraphics[width=0.35\textwidth]{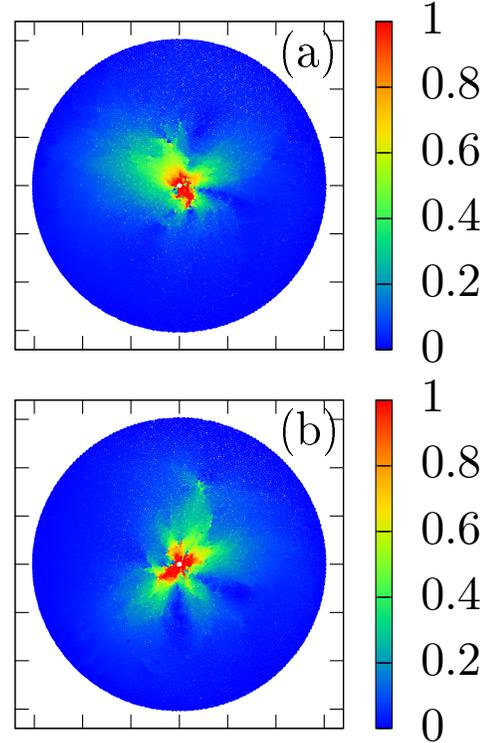}
	\caption{Examples of quasi-elastic responses associated with inflation of the inner disk
		at $\rin$ for the inverse power-law potential. These examples correspond with panels (a) and (b) of Fig.~\ref{radialIPL1}. The parameters associated with these examples are $T_m=0.4$ and 0.3 respectively, $d_0=0.15$ and 0.18, $\rin=0.72$ and 0.8, and $\rout =76$ for both. }
	\label{dispIPL1}
\end{figure}
%%%%%%%%%%%%%%%%%%%%%%%%%%%%%%%%%%%%%%%%%%%%%%%%%%%%%%%%%%%
responses that deviate significantly from the predictions of classical elasticity theory \cite{21LMMPRS,21MMPRSZ,22BMP}. The basic screening theory  was presented in a recent paper \cite{21LMMPRS}. Attention was given to the plastic responses, which typically appear as quadrupolar (Eshelby-like) irreversible responses \cite{54Esh,99ML,06ML}. When the density of these quadrupoles is low, they act only to renormalize the elastic moduli, but they do not change the form of the theory. This is reminiscent the role of dipoles in dielectrics, where the dielectric constant is dressed, but the structure of electrostatic theory remains intact \cite{84LL}. On the other hand, when the density of quadrupoles is high, the gradients of their density cannot be
ignored, and these are acting effectively as dipoles, analogous to dislocations in crystalline matter. Dipole-dipole and dipole-displacement interaction become crucial, and these change the structure of the screening-theory and the resulting mechanical responses.  A short summary of the theory is presented in Sect.~\ref{review}.

%%%%%%%%%%%%%%%%%%%%%%%%%%%%%%%%%%%%%%%%%%%%%%%%%
\begin{figure}[tb!]
	\includegraphics[width=0.29\textwidth]{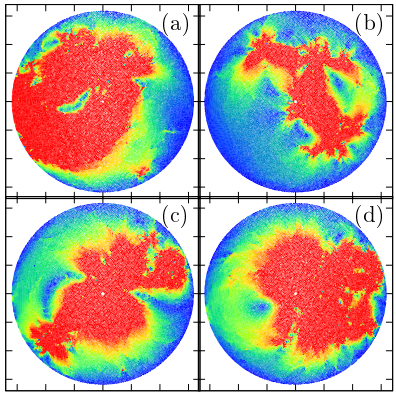}
	\caption{Examples of ``anomalous" displacement fields associated with an inflation 
		of the inner disk at $\rin$. The color codes is the same as in the previous figure. The examples shown here correspond to panels (a)-(d) in Fig.~\ref{radialIPL2}. }
	\label{dispIPL2}
\end{figure}
%%%%%%%%%%%%%%%%%%%%%%%%%%%%%%%%%%%%

%%%%%%%%%%%%%%%%%%%%%%%%%%%%%%%%%%%%%%%%%%%%%%%%%%%%%%%%%%%
\begin{figure}[htb!]
	\includegraphics[width=0.28\textwidth]{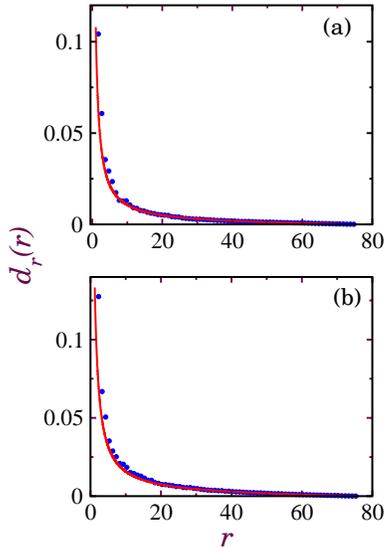}
	\caption{Typical quasi-elastic angle-averaged displacement field. }
	\label{radialIPL1}
\end{figure}

%%%%%%%%%%%%%%%%%%%%%%%%%%%%%%%%%%%%%%%%%%%%%%%%%%%%%%%%%%

The screening theory of amorphous solids is based on a Lagrangian constructed from symmetry arguments, without referring to microscopic interactions.  Nevertheless, until the present time the predictions of the novel theory were examined only in the
context of granular matter, in both simulations and experiments \cite{21LMMPRS,21MMPRSZ}. Due to the general importance of the issues involved, it is advisable to test and examine the relevance of the theory to as many different types of amorphous solids as possible. The aim of this paper is to do so with classical models of glass formers, especially those that have been used extensively to shed light on the physics of glasses.  
These models employ point particles with binary interactions. We examine here two classes or models, one with inverse power-law interactions, and thus having only repulsive forces, and one with Lennard-Jones interactions, having both repulsive and an attractive forces depending on the distance between the particles. We find results that are in accord with the novel theory of screened elasticity, where displacement fields can deviate strongly from the expectations of elasticity theory. 

In Sect.~\ref{review} we present a short summary of the theory. Rather than discussing
the theory in complete generality we concentrate on glasses in radial geometry where
the equations and the predictions take a particularly simple form. This will help us
to discuss the results of numerical simulations that are presented in Sects.~\ref{Inverse} and \ref{LJ}. Models with inverse power-law interactions are discussed in Sect.~\ref{Inverse} whereas Lennard-Jones potentials are the subject of Sect.~\ref{LJ}. In Sect.~\ref{corr} we link the degree of deviation from quasi-elastic response to the amount of energy dissipated in plastic events.  In Sect.~\ref{summary} we offer a summary and discussion.

%%%%%%%%%%%%%%%%%%%%%%%%%%%%%%%%%%%%%%%%%%
\begin{figure}
	\includegraphics[width=0.40\textwidth]{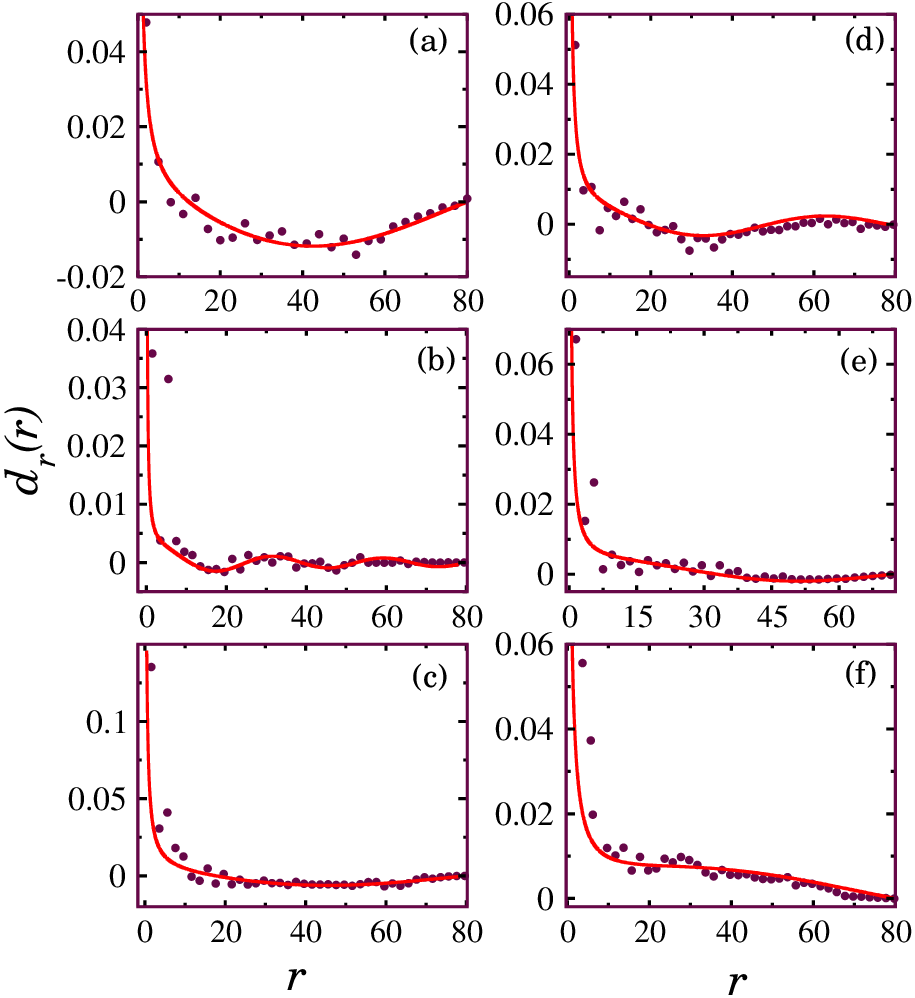}
	\caption{Typical anomalous angle-averaged displacement fields. The parameters of the simulation and the value of the ``anmoaly parameter" $\kappa$ are presented in table~\ref{table1}.}
	\label{radialIPL2}
\end{figure}
%%%%%%%%%%%%%%%%%%%%%%%%%%%%%%%%%%%%%%%%%%%%%%%%%%%%%%%%%%%%%%%%%%%

\section{Brief review of the theory} 
\label{review}

To introduce the theory in its simplest form
we focus here on 2-dimensional systems with radial geometry. We construct below amorphous configurations of $N$ point particles in an annulus, confined between two disks, and
examine the displacement fields that result from an inflation of the inner disk. This geometry was found very useful before in the context of assemblies of frictionless and frictional disk.  
Details of these simulations and experiments can be found in Refs.~\cite{21LMMPRS,21MMPRSZ}. We thus consider an annulus of radii $\rin$ and $\rout$, $\rin\ll \rout$, with an imposed displacement $\mathbf{d}(\rin) = d_0 \hat{r}$ and $\mathbf{d}(\rout) = 0$. The polar symmetry of the problem implies that $\mathbf{d}(r) =d_r(r) \hat{r}$. Whenever normal elasticity applies, this radial displacement field satisfies the equation
\begin{equation}
	\Delta {\mathbf{d}}=	d_r'' +\frac{d_r'}{r}  -\frac{d_r}{r^2}=0 \ .
	\label{usual}
\end{equation}
This differential equation is readily solved, and to satisfy the boundary conditions we write
\begin{equation}
	d^{\rm el}_r(r)=d_0 \frac{r^2 - \rout^2}{\rin^2 - \rout^2}\frac{\rin}{r} \ .
	\label{renelas}
\end{equation}
One should note that the solution (\ref{renelas}) is a positive monotonically decreasing function of $r$ as is expected in standard elasticity theory. It was shown \cite{21LMMPRS} that this form of the displacement field is expected to
remain valid also when there exists a low (or uniform) density of screening plastic events.  We  refer to this situation as weekly screened, or ``quasi-elastic" since the elastic moduli tend to get renormalized without changing the qualitative elastic response.

When the density of plastic events gains sizeable gradients, these act as dipole sources \cite{21LMMPRS} and the theory changes qualitatively. The screening caused by dipoles modify Eq.~(\ref{renelas}) to read
\begin{equation}
	d_r'' +\frac{d_r'}{r}  -\frac{d_r}{r^2}= -\kappa^2 d_r  \,
	\label{new}
\end{equation}
with $\kappa$ being an emergent constant that is evaluated and discussed below. We refer to $\kappa$ as the ``screening parameter". Eq.~(\ref{new}) is equivalent to the Bessel equation whose solution, 
satisfying $d_r(r_{\rm in})=d_0$, $d_r(r_{\rm out})=0$, reads
\begin{equation}
	d_r(r)  = d_0 \frac{ Y_1(r \, \kappa ) J_1(r_\text{out} \kappa )-J_1(r \, \kappa ) Y_1(r_\text{out} \kappa )}{Y_1( r_\text{in} \kappa ) J_1(r_\text{out} \kappa )-J_1(r_\text{in} \kappa ) Y_1(r_\text{out} \kappa )} \ .
	\label{amazing}
\end{equation}
Here $J_1$ and $Y_1$ are the Bessel functions of the first and second kind respectively. For very small values of $\kappa$ the solution reduces back to \eqref{renelas}. Importantly, depending on the precise value of $\kappa$, Eq.~(\ref{amazing}) may be non monotonic, negative, and even oscillatory. We refer to situations that agree
with the solution Eq.~(\ref{amazing}) and are either non-monotonic or oscillating as ``anomalous elasticity". In the rest of this paper we demonstrate the relevance of anomalous elasticity to the mechanical responses of classical glass formers.

\section{Glass formers with inverse power-law forces}
\label{Inverse}

\subsection{Construction of the simulation}

To conform with the theory described in Sect.~\ref{review} we construct an annulus
with two rigid walls, with an inner radius $\rin$ and outer radius $\rout$. The annulus is then filled up with $N$ point particles put in random positions in the area $A=\pi (\rout^2-\rin^2)$. The number of particles is chosen such that the density of the equilibrated glass (as described below) is a chosen value $\rho=N/A$. In this section
we use  a standard poly-dispersed model of $N$ particles of mass $m=1$ \cite{17NBC,19BFFSS}. The binary interactions are
\begin{eqnarray}
	&&\phi(r_{ij}) = \epsilon\left(\frac{\sigma_{ij}}{r_{ij}}\right)^{12} +C_0 +C_2\left(\frac{r_{ij}}{\sigma_{ij}}\right)^2+C_4\left(\frac{r_{ij}}{\sigma_{ij}}\right)^4\nonumber \\&& \epsilon\!=1, C_0\!=\!-1.92415, C_2\!=\!2.11106, C_4\!=\!-0.591097 \ . \label{IPL}
\end{eqnarray}
The unit of energy is $\epsilon$ and Boltzmann's constant is unity.
The interaction length was drawn from a probability distribution $P(\sigma)
\sim 1/\sigma^3$ in a range between $\sigma_{\rm min}$ and $\sigma_{\rm max}$ such that the mean $\bar \sigma=1$:
\begin{eqnarray}
	&&\sigma_{ij} =\frac{\sigma_i+\sigma_j}{2}\Big[1-0.2\Big|\sigma_i-\sigma_j\Big|\Big],\nonumber\\ 
	&&\sigma_{\rm max} =1.45/0.9\ , \sigma_{\rm min}=\sigma_{\rm max}/2.219 \ .
\end{eqnarray}
The units of mass and length are $m$ and $\bar \sigma$ (the average $\sigma$). The parameters are chosen to avoid crystallization. The system is thermalized at some ``mother temperature" $T_m$ using Swap Monte Carlo and then 
cooled down to $T=0$ using conjugate gradient methods. The interaction between the point particles and the two walls are of the same form Eq.~(\ref{IPL}), where $r_{ij}$ and $\sigma_{ij}$ are replaced by the distance to the wall and by $\sigma_i$. 

Once the system is mechanically equilibrated with the total force on each particle smaller than $10^{-8}$, we inflate the inner radius $\rin$ by some percentage as
reported below. After inflation we equilibrate the system again by the conjugate gradients, and then measure the displacement field $\B d$, comparing the two equilibrated configuration before and after inflation. To compare with the theory we then compute the angle-averaged radial component of the displacement field $d_r(r)$.

\subsection{results of simulations}

In Fig.~\ref{dispIPL1} and \ref{dispIPL2} we show typical results for the total displacement fields after inflation. The color code designates the absolute magnitude of the vector displacement field. We note that sometimes the displacement fields are concentrated around the origin, and sometimes they are spread out in the system, implying the importance of quadrupole gradients. To test the theory we compare the angle-averaged radial displacement with the predicted fields. We should add here that there can be also higher order responses that are not captured by angle averaging. The discussion of higher order modes is deferred to a followup paper. 

The first type of responses (cf. Fig.~\ref{dispIPL1})  agrees  with the quasi-elastic result Eq.~(\ref{renelas}); in the other case (Fig.~\ref{dispIPL2}) the measured radial components conform with 
Eq.~(\ref{amazing}). Below we rationalize these finding, relating the degree of anomaly in the response with the amount of energy lost to plastic events. 
%%%%%%%%%%%%%%%%%%%%%%%%%%%%%%%%%%%%%%%%%%%%%%
Typical angle-averaged displacement are shown in Figs.~\ref{radialIPL1} and \ref{radialIPL2}. 
%%%%%%%%%%%%%%%%%%%%%%%%%%%%%%%%%%%%%%%%%%%%%%%%%%%% 
\begin{table}
	\begin{center}
		\begin{tabular}{c|c|c|c|c|c}
				panel & $T_m$ & $\rin$&$\rout$&$d_0$&$\kappa$ \\
			\hline
			(a) & 2.0 & 0.65 & 80&0.101&0.051  \\ \hline
			(b)& 0.2 & 0.5 & 80 &0.05 & 0.229 \\ \hline
			(c)& 0.3 & 0.65 & 80 &0.10 & 0.054 \\ \hline
			(d)& 0.5 & 0.5 & 78 &0.09 & 0.109 \\ \hline
			(e)& 0.4 & 0.5 & 72 &0.07 & 0.082 \\ \hline
			(f)& 0.2 & 0.7 & 80 &0.1 & 0.042 \\ \hline
			\label{table}
		\end{tabular}
	\end{center}
	\caption{The parameters associated with the radial component of the displacement field shown in Fig.~\ref{radialIPL2}.}
	\label{table1}
\end{table}
%%%%%%%%%%%%%%%%%%%%%%%%%%%%%%%%%%%%%%%%%%%%%%%%%%%%%%
While Eq.~(\ref{renelas}) provides the correct functional form that fits the measured displacement fields in Fig.~\ref{radialIPL1}, all the examples shown in Fig.~\ref{radialIPL2}
are in accord with the solutions (\ref{amazing}) with different values of $d_0$ and $\kappa$.
We stress that the functional fits in Figs.~\ref{radialIPL1} and \ref{radialIPL2} are performed using the actual values of $\rin$ and $\rout$ and $d_0$ and only $\kappa$ is fitted. We observe that the functional forms in Fig.~\ref{radialIPL2} vary considerably,
with one broad minimum, one or more oscillations, etc. None of these examples can be fitted
to the functional form (\ref{renelas}). The parameters of the simulations
and of values of the screening parameter $\kappa$ are collected in 
table~\ref{table1}, in correspondence with the panels of Fig.~\ref{radialIPL2}.  

\section{Glass formers with Lennard-Jones potential}
\label{LJ}

\subsection{Construction of the simulation}

In this section, we study a two-dimensional poly-dispersed model
of point particles having equal mass $m=1$, with interaction given by shifted and smoothed Lennard-Jones (LJ) potentials, $u_{}(r)$,
\begin{equation}
	u_{ij}(r) = \begin{cases} u^{LJ}_{ij}+A_{ij} +B_{ij}r+C_{ij}r^2, & \mbox{if } r \leq R^{cut}_{ij}
		\\ 0, & \mbox{if } r > R^{cut}_{ij}, \end{cases}
	\label{Usmooth}
\end{equation}
where
\begin{equation}
	u^{LJ}_{ij} = 4\epsilon_{ij}\left[\left(\frac{\sigma_{ij}}{r}\right)^{12} - \left(\frac{\sigma_{ij}}{r}\right)^6\right].
	\label{ULJ}
\end{equation}
The smoothing  of potentials in Eq.~(\ref{Usmooth}) is such that they vainsh with
two zero derivatives at distances $R^{cut}_{ij} = 2.5\sigma_{ij}$ \cite{09LP}. The interaction lengths $\sigma_{ij}$ are chosen as in Sect.~\ref{Inverse} from the probability distribution $P(\sigma)$.  
The parameters for smoothing the LJ potentials in Eq. (\ref{Usmooth}) and
for $i$ and $j$ particle interactions in Eq.(\ref{ULJ}) are as follows: $A_{ij}=0.4526\epsilon_{ij}$, $B_{ij}=-0.3100\epsilon_{ij}/\sigma{ij}$, $C_{ij}=0.0542\epsilon_{ij}/\sigma{ij}^2$.
The reduced units for mass, length, energy and time have been taken as $m$,
$\bar \sigma$, $\epsilon_{ij}=1$ and $\bar\sigma\sqrt{m/\epsilon_{ij}}$ respectively.

The system preparation protocol is identical to the one described above for the power-law interaction. The inflation of the central disk and the calculation of the displacement field follows accordingly. 

\subsection{Results of simulations}

\begin{figure}
	\includegraphics[width=0.36\textwidth]{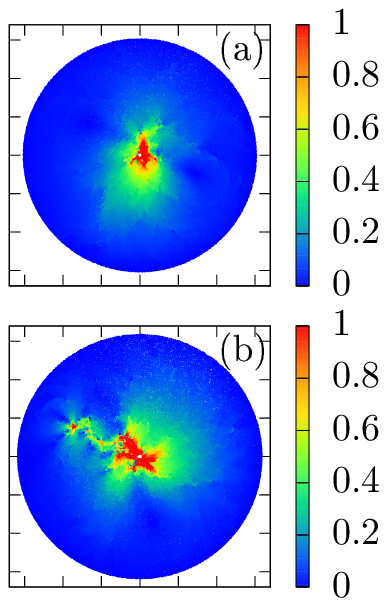}
	\caption{Examples of quasi-elastic responses associated with inflation of the inner disk
		at $\rin$ for the Lennard-Jones potential. These examples correspond with panels (a) and (b) of Fig.~\ref{radialIPL1}. The parameters associated with these examples are $T_m=0.9$ and 0.5 respectively, $d_0=0.1$ and 0.15, $\rin=0.6$ and 0.75, and $\rout =77$ and 80. }
	\label{LJQ}
\end{figure}

\begin{figure}
	\includegraphics[width=0.30\textwidth]{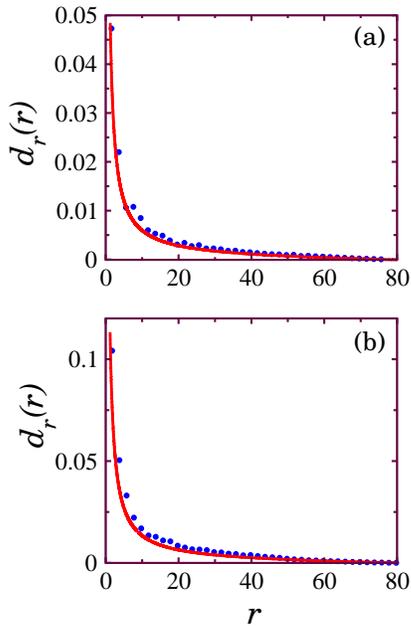}
	\caption{Typical quasi-elastic angle-averaged displacement field for configurations shown in Fig.~\ref{LJQ}. }
	\label{LJan}
\end{figure}

As for the inverse power-law potential, also configurations with Lennard-Jones potential can exhibit quasi-elastic and anomalous displacement fields. Example of each type of response are shown in Fig.~\ref{LJQ} and \ref{LJan}.
%%%%%%%%%%%%%%%%%%%%%%%%%%%%%%%%%%%%%%%%%%%%%%%%%%%%%%%%%%%
%%%%%%%%%%%%%%%%%%%%%%%%%%%%%%%%%%%%%%%%%%%%%%%%%%%%%%%%%%%
In Fig.~\ref{LJan} we show the angle-averaged displacement field associated with the examples shown in Fig.~\ref{LJQ}.
%%%%%%%%%%%%%%%%%%%%%%%%%%%%%%%%%%%%%%%%%%%%%%%%%%%%%%%%%%%
%%%%%%%%%%%%%%%%%%%%%%%%%%%%%%%%%%%%%%%%%%%%%%%%%%%%%%%%%%%

\begin{figure}
	\includegraphics[width=0.38\textwidth]{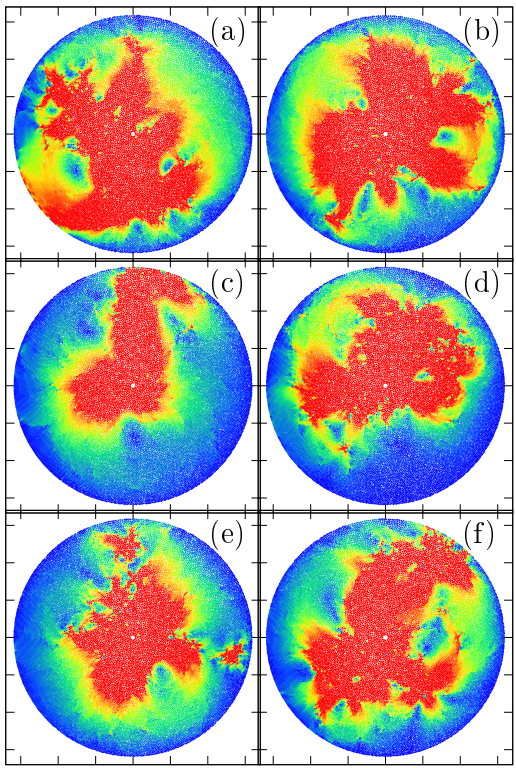}
	\caption{Examples of ``anomalous" displacement fields associated with an inflation 
		of the inner disk at $\rin$. The color code here is identical to the one used in Fig.~\ref{LJQ}. The examples shown here correspond to panels (a)-(d) in Fig.~\ref{radialLJ2}. }
	\label{LJmaps}
\end{figure}

Anomalous responses for the Lennard-Jones case are shown in Fig.~\ref{LJmaps}
%%%%%%%%%%%%%%%%%%%%%%%%%%%%%%%%%%%%%%%%%%%%%%
%%%%%%%%%%%%%%%%%%%%%%%%%%%%%%%%%%%%
%%%%%%%%%%%%%%%%%%%%%%%%%%%%%%%%%%%%%%%%%%%%%%%%%%%%%%%%%%%
\begin{figure}
	\includegraphics[width=0.45\textwidth]{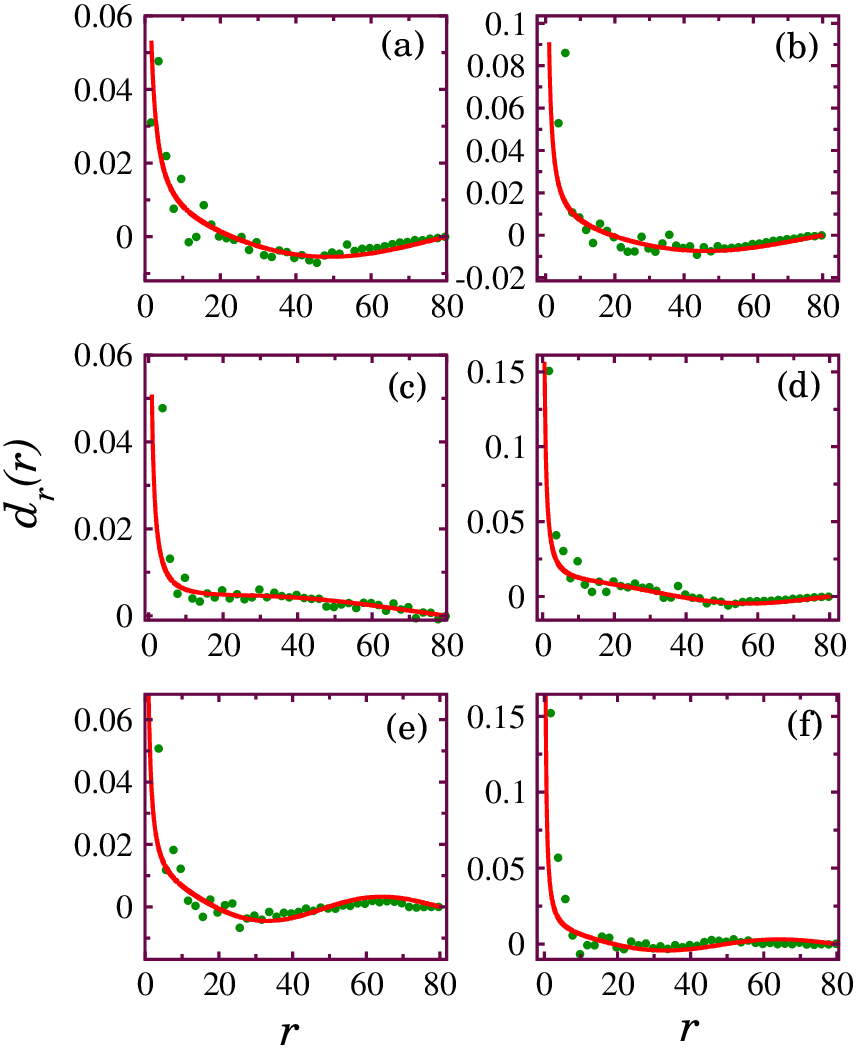}
	\caption{Typical anomalous angle-averaged displacement fields. The parameters of the simulation and the value of the ``screening parameter" $\kappa$ are presented in table~\ref{table2}.}
	\label{radialLJ2}
\end{figure}
%%%%%%%%%%%%%%%%%%%%%%%%%%%%%%%%%%%%%%%%%%%%%%%%%%%%%%%%%%%%%%%%%%%
%%%%%%%%%%%%%%%%%%%%%%%%%%%%%%%%%%%%%%%%%%%%%%%%%%%% 
\begin{table}
	\begin{center}
		\begin{tabular}{c|c|c|c|c|c}
			panel & $T_m$ & $\rin$&$\rout$&$d_0$&$\kappa$ \\
			\hline
			(a) & 1.0 & 0.6 & 80&0.15&0.058  \\ \hline
			(b)& 0.9 & 0.7 & 80 &0.13 & 0.055 \\ \hline
			(c)& 1.0 & 0.72 & 80 &0.06 & 0.042 \\ \hline
			(d)& 1.0 & 0.75 &80 &0.11 & 0.076 \\ \hline
			(e)& 0.9 & 0.63 & 80 &0.1 & 0.105 \\ \hline
			(f)& 0.9 & 0.75 & 80 &0.12 & 0.11 \\ \hline
			\label{table}
		\end{tabular}
	\end{center}
	\caption{The parameters associated with the radial component of the displacement field shown in Fig.~\ref{radialLJ2}.}
	\label{table2}
\end{table}
%%%%%%%%%%%%%%%%%%%%%%%%%%%%%%%%%%%%%%%%%%%%%%%%%%%%%%
The associated angle-averaged displacement fields are presented in Fig.~\ref{radialLJ2}, with corresponding panels. The parameters of the simulations and of the values of $\kappa$ for each panel are presented in table~\ref{table2}.

\section{Correlation between plasticity and screening}
\label{corr}
\begin{figure}
	\includegraphics[width=0.35\textwidth]{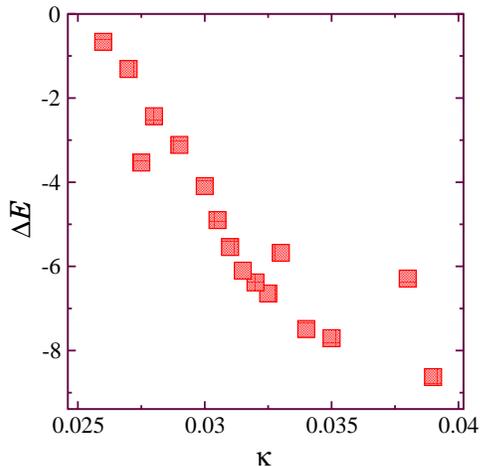}
	\caption{The correlation between the amount of energy dissipation due to the plasticity and the numerical value of the screening parameter $\kappa$}
	\label{kapE}
\end{figure}

At this point we should raise the obvious question: when do we expect quasi-elastic responses, and when is screening sufficient to alter the situation and cause anomalous responses. Equivalently we can ask what determines the value of the screening parameter $\kappa$. Since we expect screened elasticity when the density of quadrupoles is high, it is tempting to correlated the value of $\kappa$ with the amount of energy lost to dissipation. To quantify the amount of dissipation we note that before inflation of the central disk the total energy of the system is given as $E_0$. After inflation and
equilibration the energy is $E$, and the energy difference is $\Delta E\equiv E-E_0$.  Obviously, if there is no plasticity involved in the inflation and equilibration, $\Delta E$ is maximal and positive, and $\kappa=0$. Plastic dissipation is consistent with negative $\Delta E$. 
Examining our displacement fields for a given configuration with different $d_0$, and the fitted values of $\kappa$, we plot $\Delta E$ vs. $\kappa$ and find that (up to expected
fluctuations) there is a clear functional dependence between the two, cf. Fig.~\ref{kapE}.
The higher is the dissipation (as seen by a more negative $\Delta E$),
the higher is the value of $\kappa$
%%%%%%%%%%%%%%%%%%%%%%%%%%%%%%%%%%%%%%%%%%%%%%%%%%%%%%%%%%%
%%%%%%%%%%%%%%%%%%%%%%%%%%%%%%%%%%%%%%%%%%%%%%%%%%%

One can introduce a direct measure of the degree of deviation from the quasi-elastic response Eq.~(\ref{renelas}), using the norm of deviation $M$: 
\begin{equation}
	M \equiv \frac{1}{r_{out}-r_{in}}\int_{r_\text{in}}^{r_\text{out}} \Big|\frac{	d^{\rm el}_r(r) -d_r(r)}{	d^{\rm el}_r(r)}\Big|	\ .
	\label{defM}
\end{equation}
When the actual angle averaged displacement field $d_r(r)$ equals the quasi-elastic one, $M=0$. When $M$ differs from zero, it means that plastic screening is at play.

\begin{figure}[htb!]
	\includegraphics[width=0.35\textwidth]{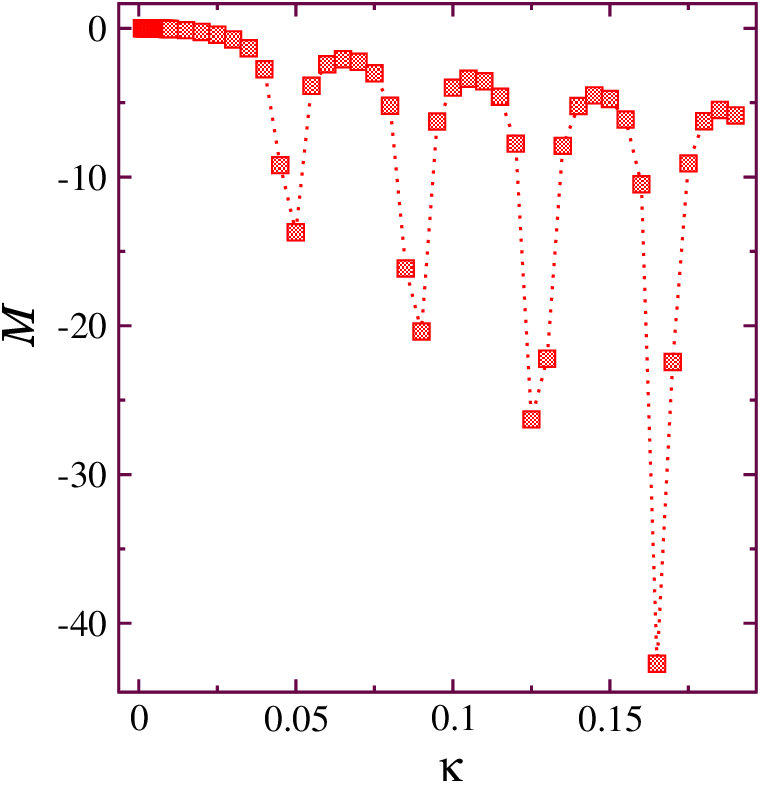}
	\caption{The dependence of $M$ on $\kappa$ for a L-J system with given $r_\text{in}$, $r_\text{out}$ and $d_0$}
	\label{Mkappa}
\end{figure}

The dependence of $M$ on the screening parameter $\kappa$ is not necessarily monotonic. To exemplify this we consider anomalous 
displacement fields for a system with given $r_\text{in}$, $r_\text{out}$ and $d_0$, as a function of $\kappa$. This functional dependence is shown in Fig.~\ref{Mkappa}.
%%%%%%%%%%%%%%%%%%%%%%%%%%%%%%%%%%%%%%%%%%%%%%%%%%%%%%%%%%%
%%%%%%%%%%%%%%%%%%%%%%%%%%%%%%%%%%%%%%%%%%%%%%%%%%%%
We observe an oscillatory dependence with increasing negative values for
$M$ as $\kappa$ increases.

\begin{figure}
	\includegraphics[width=0.35\textwidth]{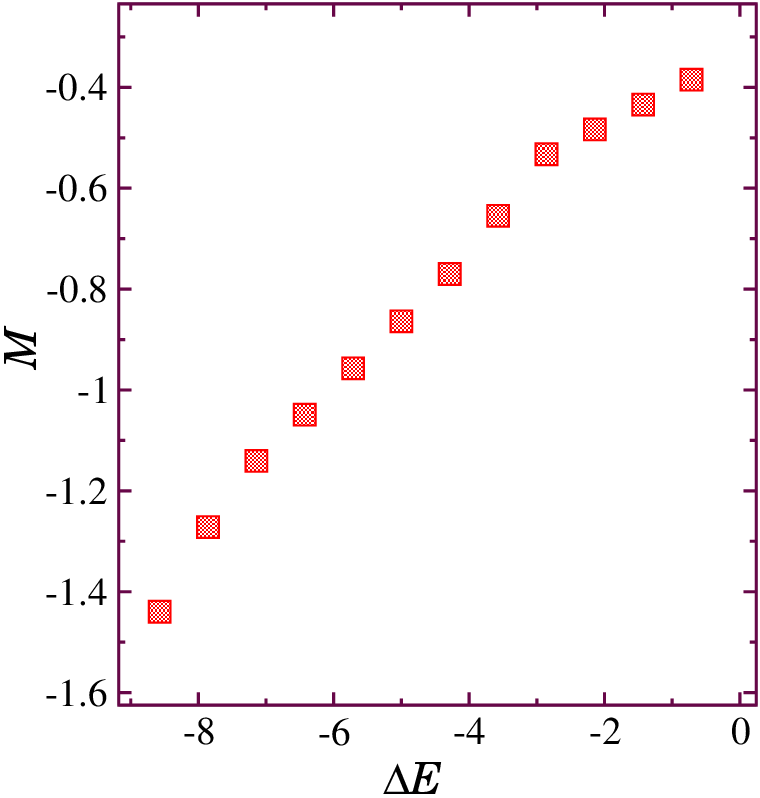}
	\caption{An example of the dependence of $M$ on $\Delta E$ for a L-J system prepared from $T_m=1$, with given $r_\text{in}=0.72$, $r_\text{out}=80$ and variable $d_0$ in the range $[10\%-40\%]$. }
	\label{ME}
\end{figure}

On the other hand, for the same protocol used in Fig.~\ref{kapE}, the observed value of $M$ is monotonic in the amount
of plastic dissipation. To see this one needs to consider again a given
initial configuration with a given  $r_\text{in}$ and $r_\text{out}$,
for different values of $d_0$.
We find that $M$ is a smooth function of $\Delta E$ as is exemplified in Fig.~\ref{ME}.
%%%%%%%%%%%%%%%%%%%%%%%%%%%%%%%%%%%%%%%%%%%%%%%%%%%%%%%%%%%

%%%%%%%%%%%%%%%%%%%%%%%%%%%%%%%%%%%%%%%%%%%%%%%%%%%%
With $\Delta E$ progressively negative, meaning more plastic dissipation, also $M$ becomes progressively negative, indicating larger deviation from the quasi-elastic response. 
%%%%%%%%%%%%%%

It should be stressed that the monotonic dependence shown in Fig.~\ref{ME} is observed by fixing a given glassy configuration. Different configurations suffer from the usual sample-to-sample fluctuations that typify amorphous solids. In other words, different configurations with the same $r_\text{in}$, $r_\text{out}$, and  $d_0$ can show different values of $M$ and $\kappa$. Only for a given configuration one can present a monotonic dependence of the degree of deviation from elasticity and plastic dissipation. 

\section{summary and discussion}
\label{summary}

The results of this paper extend the degree of genericity of anomalous elasticity through screening by plastic quadrupoles and dipoles. In previous papers the discussion was limited to granular material with and without friction. In the present paper we examined classical glass formers of point particles with purely repulsive interactions and with L-J force laws that combine repulsion and attraction. Generally speaking all these examples of amorphous solids support the assessment that anomalous elasticity is generic and needs to be taken into account in studying the mechanical responses of amorphous solids to non-uniform strains.

A new ingredient in the present paper is the direct relation of the degree of deviation of the radial displacement from the quasi-elastic counterpart. Besides the obvious meaning of the screening parameter $\kappa$, we proposed a measure of this deviation, the quantity $M$ of Eq.~(\ref{defM}), and showed that it has a functional dependence on the amount of plastic dissipation, cf.~Fig.~\ref{ME}. 

It should be stressed that the screening theory that was presented so far is linear, in the sense that the Lagrangian taken to represent the state of the system is developed up to quadratic order in the quadrupolar and dipolar fields.  In future work we will develop the nonlinear theory as well,
to be able to describe also responses that occur at larger values
of strains, including shear bands, steady plastic flow and material
failure. Another direction for the immediate future is the extension
of the analysis to 3-dimensional systems, in which the nature of screening may be different from the presently understood 2-dimensional examples. Finally, the nonlinear extension of the theory will allow us to investigate the excitation of higher order angular modes by the  isotropic expansion of a single particle.

{\bf Acknowledgments}: This work had been supported in part by a collaborative grant of the Israel Science Foundation and China, grant \# 714173. Additional funds were presented by the Minerva Foundation and the Minerva Center for ``Aging, from physical materials to human tissues" at the Weizmann Institute.  MM acknowledges support from the Israel Science Foundation (grant No. 1441/19).

\bibliography{glassformers}

\end{document}